\documentclass[aps,prb,twocolumn,
10pt,%longbibliography,%preprint,
groupedaddress]{revtex4-1}
\usepackage{graphicx,bm}
\def\be{\begin{equation}} 
\def\ee{\end{equation}} 
\def\ba{\begin{eqnarray}} 
\def\ea{\end{eqnarray}} 
\def\bc{\begin{center}} 
\def\ec{\end{center}} 
 
\def\p{\partial} 
\def\di{\textrm{div\,}}
\def\ditwo{\textrm{div}_2\,}
\def\rot{\textrm{rot\,}}

\begin{document}

\title{Equations of macroscopic electrodynamics for two-dimensional crystals}

\author{S. A. Mikhailov}
\email[]{sergey.mikhailov@physik.uni-augsburg.de}
\affiliation{Institute of Physics, University of Augsburg, D-86135 Augsburg, Germany}

\date{\today}

\begin{abstract}
The electrodynamics of two-dimensional (2D) dielectric and conducting layers cannot be described by such three-dimensional macroscopic quantities as the dielectric constant $\epsilon$  or the refractive index $n$. By means of the proper averaging of the microscopic Maxwell equations we derive general macroscopic electrodynamic equations for 2D crystals and discuss some of their consequences.
\end{abstract}

%\pacs{78.67.-n, 78.67.Wj}

\maketitle

Equations of the \textit{macroscopic} electrodynamics of bulk (three-dimensional, 3D) materials are derived from the \textit{microscopic} Maxwell equations 
\be 
\begin{array}{ll}
\di \textbf{e}=4\pi \rho,&\hspace{5mm} \di \textbf{h}=0, \\
\rot\textbf{e}=-\frac 1c\frac{\p \textbf{h}}{\p t},&\hspace{5mm} \rot\textbf{h}=\frac 1c\frac{\p \textbf{e}}{\p t} +\frac{4\pi }c\textbf{j}\\
\end{array}
\ee 
by their averaging over ``physically infinitesimal'' volume elements \cite{Landau8}, i.e. over dimensions small as compared to the macroscopic scales, such as the wavelength of radiation and the sample dimensions, but large as compared to the inter-atomic distances. Here $\textbf{e}$, $\textbf{h}$, $\rho$ and $\textbf{j}$ are the \textit{microscopic} electric and magnetic fields, charge and current densities. As a result of the averaging one defines the \textit{macroscopic} electric field $\textbf{E}=\bar \textbf{e}$, the polarization vector $\textbf{P}=\chi \textbf{E}$, which has the meaning of the dipole moment of a \textit{volume} element, the electric induction $\textbf{D}=\textbf{E}+4\pi \textbf{P}$, the dielectric susceptibility $\chi$, the dielectric constant $\epsilon=1+4\pi\chi$ and other macroscopic quantities. The dimensionless quantities $\chi$ and $\epsilon$ (we use the more physical Gaussian system of units), which can in general be functions of the frequency $\omega$ and the wave-vector $\textbf{q}$ of the electromagnetic field, fully determine the linear response of the medium to the electromagnetic field. In the nonlinear optics one defines\cite{Boyd08} the higher order susceptibilities, e.g., the third order tensor $\chi^{(3)}_{ijkl}$ which determines the cubic term in the Taylor expansion of the function $\textbf{P}(\textbf{E})$.

The discovery of graphene \cite{Novoselov04,Novoselov05,Zhang05} -- a one-atom-thick layer of carbon atoms -- triggered great interest to this, as well as other two-dimensional (2D) materials \cite{Novoselov05Nat}, both metallic (graphene), dielectric (e.g. BN) and semiconducting (e.g. MoS$_2$). The notions of the dielectric susceptibility $\chi$, the dielectric constant $\epsilon$ and the refractive index $n=\sqrt{\epsilon}$ are no longer applicable to these materials since the averaging of microscopic fields over the physically infinitesimal volume elements is impossible in the direction perpendicular to the 2D layer. However, by analogy with 3D materials, many authors continue to characterize the electrodynamic response of graphene and other 2D crystals, especially their nonlinear properties, by 3D quantities such as the third susceptibility $\chi^{(3)}$ or the nonlinear refractive index $n_2$. 
A basic inadequacy of such an approach is evident: the refractive index $n$ characterizes the change of the phase velocity of the wave \textit{propagating inside the material}, but it makes no sense to talk about the propagation of waves inside a one-atom-thick layer. 

Thus a fundamental question arises, how to write down the macroscopic electrodynamic equations for atomically-thin (in fact, two-dimensional) materials and which physical quantity should be used to characterize their linear and nonlinear electrodynamic and optical properties. 

In a 3D dielectric the averaging of the microscopic charge density over the physically infinitesimal {\em volume} elements leads to the definition of the polarization vector $\textbf{P}$, $ \rho \rightarrow \bar \rho  = - \di \textbf{P}$, which has the meaning of the dipole moment per unit volume \cite{Landau8} and has the dimension [$e$/cm$^2$]. In two dimensions such an averaging can be performed only over a physically infinitesimal {\em surface} element, 
\be 
\rho \rightarrow \delta(z)\bar \rho(\textbf{r}_\parallel)  = - \delta(z)\ditwo  \textbf{P}_\parallel(\textbf{r}_\parallel),
\ee
where the subscripts $\parallel$ and 2 indicate 2D vectors or operators, the vector $\textbf{P}_\parallel=(P_x,P_y)$ is the dipole moment of a \textit{surface} element (the dimension [$e$/cm]), and we consider the polarization in the direction parallel to the layer. The corresponding contribution to the polarization current (the dimension statampere/cm) is 
\be 
\textbf{j}_\parallel=\p\textbf{P}_\parallel/\p t.\label{j-P}
\ee 
In the presence of external charges $\rho_{\rm ex}(\textbf{r},t)$ and currents $\textbf{j}_{\rm ex}(\textbf{r},t)$ (which can be three-dimensional) the \textit{macroscopic} Maxwell equations for a 2D nonmagnetic medium assume the form
\ba 
\di \textbf{E}&=&-4\pi\ditwo \textbf{P}_\parallel\delta(z)+4\pi \rho_{\rm ex} ,\label{divE}\\ 
\di \textbf{H}&=&0, \\
\rot\textbf{E}&=&-\frac 1c\frac{\p \textbf{H}}{\p t},\\ 
\rot\textbf{H}&=&\frac 1c\frac{\p \textbf{E}}{\p t} +\frac{4\pi }c\frac{\p \textbf{P}_\parallel}{\p t}\delta(z)+\frac{4\pi }c\textbf{j}_{\rm ex}\label{rotH}
\ea 
Equations (\ref{divE}) -- (\ref{rotH}) should be completed by a relation between the 2D polarization vector $\textbf{P}_\parallel$ and the electric field $\textbf{E}(z=0)$. If we ignore a possible spontaneous ferroelectric polarization of a 2D crystal, predicted in Ref. \cite{Mikhailov13b}, such a relation for centrosymmetric dielectric crystals has the form 
\be 
P_{\parallel;\alpha}=\chi^{(1),\rm{2D}}_{\alpha\beta}E_{\beta}+
\chi^{(3),\rm{2D}}_{\alpha\beta\gamma\delta}E_{\beta }E_{\gamma }E_{\delta },
\label{P-E}
\ee
where the Greek indexes take only the values $\{x,y\}$ and the fields are taken at $z=0$. For conducting crystals it is more convenient to use the current-field relation
\be 
j_{\parallel;\alpha}=\sigma^{(1),\rm{2D}}_{\alpha\beta}E_{\beta}+
\sigma^{(3),\rm{2D}}_{\alpha\beta\gamma\delta}E_{\beta }E_{\gamma }E_{\delta }.
\label{j-E}
\ee
The 2D quantities $\chi^{\rm{2D}}$ and $\sigma^{\rm{2D}}$ are similar to the corresponding 3D ones but are measured in different units: for example, the first- and third-order 2D susceptibilities $\chi^{(1),\rm 2D}$ and $\chi^{(3),\rm 2D}$ are measured in cm and cm$^3$/statvolt$^2$, respectively, in contrast to the corresponding 3D quantities which are dimensionless ($\chi^{(1)}$) and measured in cm$^2$/statvolt$^2$ ($\chi^{(3)}$). In general, the (linear and nonlinear) physical quantities $\chi^{\rm{2D}}$ and $\sigma^{\rm{2D}}$ are complex functions of the frequency $\omega$ and the wave-vector $\textbf{q}_\parallel$ of the electromagnetic field. They are related to each other by formulas which can be obtained from (\ref{j-P}).

The electrodynamics of 2D materials should thus be studied using the system of equations (\ref{divE}) -- (\ref{rotH}), and their electrodynamic properties should be described by the 2D quantities $\chi^{\rm{2D}}$ or $\sigma^{\rm{2D}}$. Describing, for example, Kerr effect in graphene one should directly relate the experimentally measured quantities to the linear and nonlinear components of  $\chi^{\rm{2D}}$ or $\sigma^{\rm{2D}}$, see, e.g., Refs. \cite{Savostianova18a,Mikhailov18a}. The use of unphysical (for 2D crystals) quantities $\epsilon$, $n$ and $n_2$ is inappropriate since they cannot be rigorously defined for 2D materials. All said above refers to any material consisting of a single or a few atomic layers, including, e.g., tilted Dirac cone 2D systems \cite{Suzumura14,Jalali18a,Jalali18b} and thin films of topological insulators \cite{Jafari14}.  

Let us consider some consequences of the above equations. The linear and nonlinear dynamic conductivities of conducting graphene and carbon nanotubes have been theoretically studied in many papers,\cite{Slepyan99,Ando02,Ando06,Gusynin06a,Gusynin06b,Nilsson06,Falkovsky07a,Falkovsky07b,Mikhailov07d,Peres08,Stauber08a,Cheng14a,Cheng14b,Cheng15,Mikhailov16a,Wang16,Cheng17}. 
Electrodynamic properties of dielectric 2D materials have been discussed to a  lesser extent. Below we discuss some properties of a model dielectric 2D crystal characterized by the linear susceptibility $\chi^{(1),\rm{2D}}$. If a 2D crystal has a hexagonal lattice like graphene, the susceptibility $\chi^{(1),\rm{2D}}$ can be calculated within the tight-binding approximation assuming different on-site energies of electrons sitting on atoms of different sublattices. This model well describes the hexagonal boron nitride (BN) and was theoretically studied, for example, in Refs. \cite{Pedersen09,Jafari12} under the name of ``gapped graphene''. The spectrum of electrons and holes near the Dirac points in such a model reads
\be 
E_{l\bf k}= (-1)^l\sqrt{\Delta^2+(\hbar v_F k)^2}, \ \ l=1,2,
\ee
where $2\Delta=E_{\rm gap}$ is the energy gap and $v_F$ is the effective Fermi velocity of electrons. Assuming that the Fermi energy lies in the gap and that the temperature is small, $T\ll\Delta$, one can get the following analytical expression for the function $\chi^{(1),\rm{2D}}(\omega)$ 
\be 
\chi^{(1),\rm 2D}(\omega)=
\frac {e^2g_sg_v}{12\pi\Delta }  {\cal F}\left(\frac{\hbar |\omega|}{2\Delta}\right),
\label{chi1}
\ee
where $g_s$ and $g_v$ are the spin and valley degeneracies ($g_sg_v=4$) and 
\begin{figure}[t]
\includegraphics[width=0.98\columnwidth]{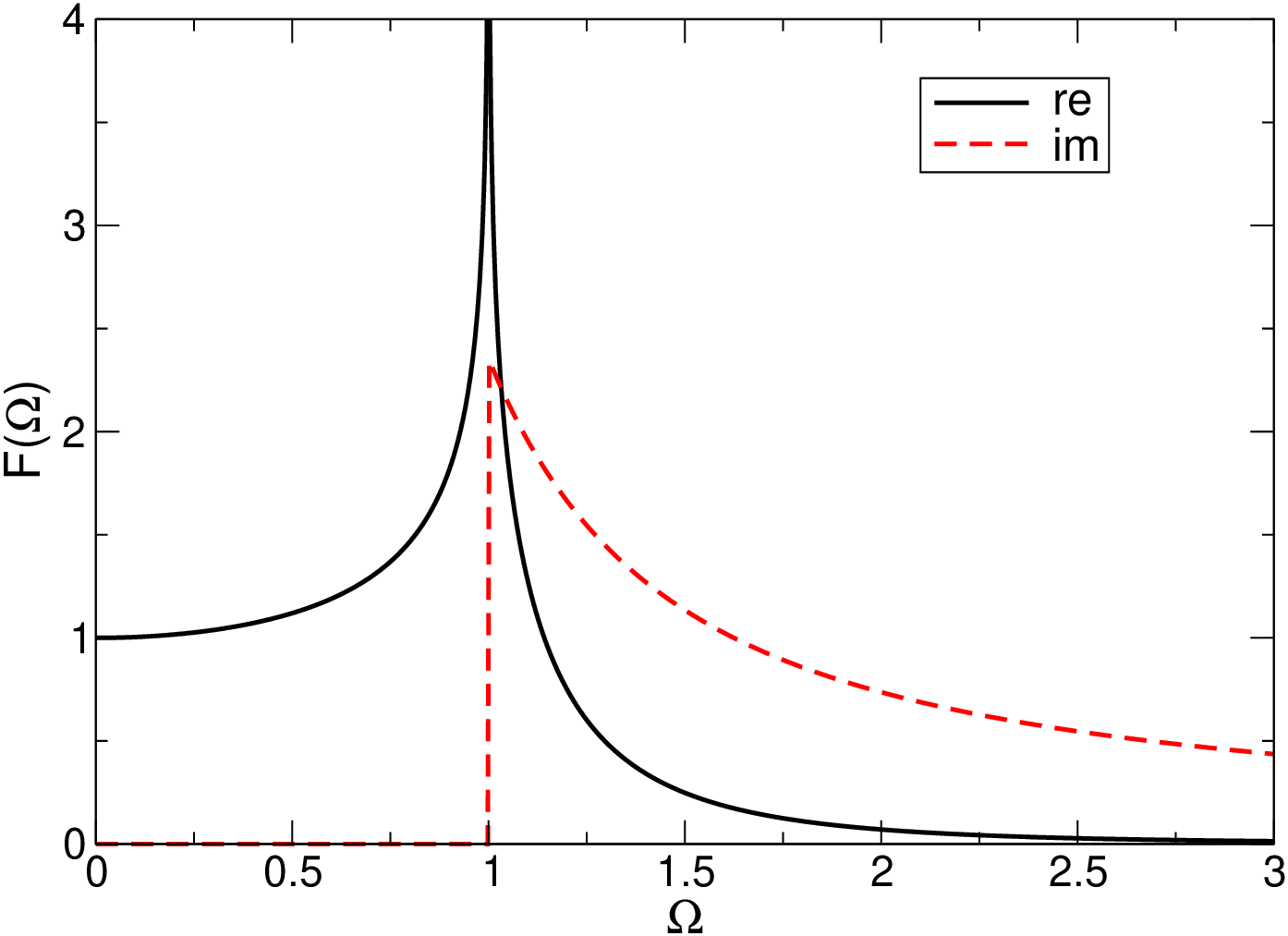}\\
\caption{\label{fig:calF} The frequency dependence of the real and imaginary parts of the function ${\cal F}(\Omega)$, defined by Eq. (\ref{calF}). }
\end{figure}
\ba 
{\cal F}\left(\Omega\right)&=&
\frac 38\left(\frac{1+\Omega^2}{|\Omega|^3}\ln\left|\frac{1+|\Omega|}{1-|\Omega|}\right| - \frac 2{\Omega^2 }\right)
\nonumber \\ &+& i\frac {3\pi}8 \Theta(|\Omega|-1)
\frac{ 1+\Omega^2}{\Omega^3}.\label{calF}
\ea
The frequency dependence of real and imaginary parts of the function ${\cal F}\left(\Omega\right)$, Eq. (\ref{calF}), is shown in Figure \ref{fig:calF}. The susceptibility has a logarithmic divergence at the frequency $\hbar|\omega|=2\Delta=E_{\rm gap}$ corresponding to the inter-band transition between the valence and conduction bands. At low frequencies $\Omega\ll 1$ the function ${\cal F}(\Omega)$ tends to unity, ${\cal F}(\Omega\to 0)=1$, so that the 2D static susceptibility is 
\be 
\chi^{(1),\rm{2D}}_{\omega\to 0}=\frac{e^2}{3\pi\Delta}.\label{chi1static}
\ee
It dramatically grows when the band gap decreases. If $\Delta=1$ eV ($E_{\rm gap}=2\Delta=2$ eV), the static susceptibility (\ref{chi1static}) equals $\chi^{(1),\rm{2D}}_{\omega\to 0}=0.153$ nm; if the gap lies in the terahertz range it is several orders of magnitude larger.

The third-order nonlinear susceptibility $\chi^{(3),\rm{2D}}$ can be calculated within the tight binding approximation, in a similar way as for graphene.\cite{Mikhailov16a} Within the Dirac Hamiltonian approach it was done in Ref. \cite{Jafari12}.

\begin{figure}[t]
\includegraphics[width=0.98\columnwidth]{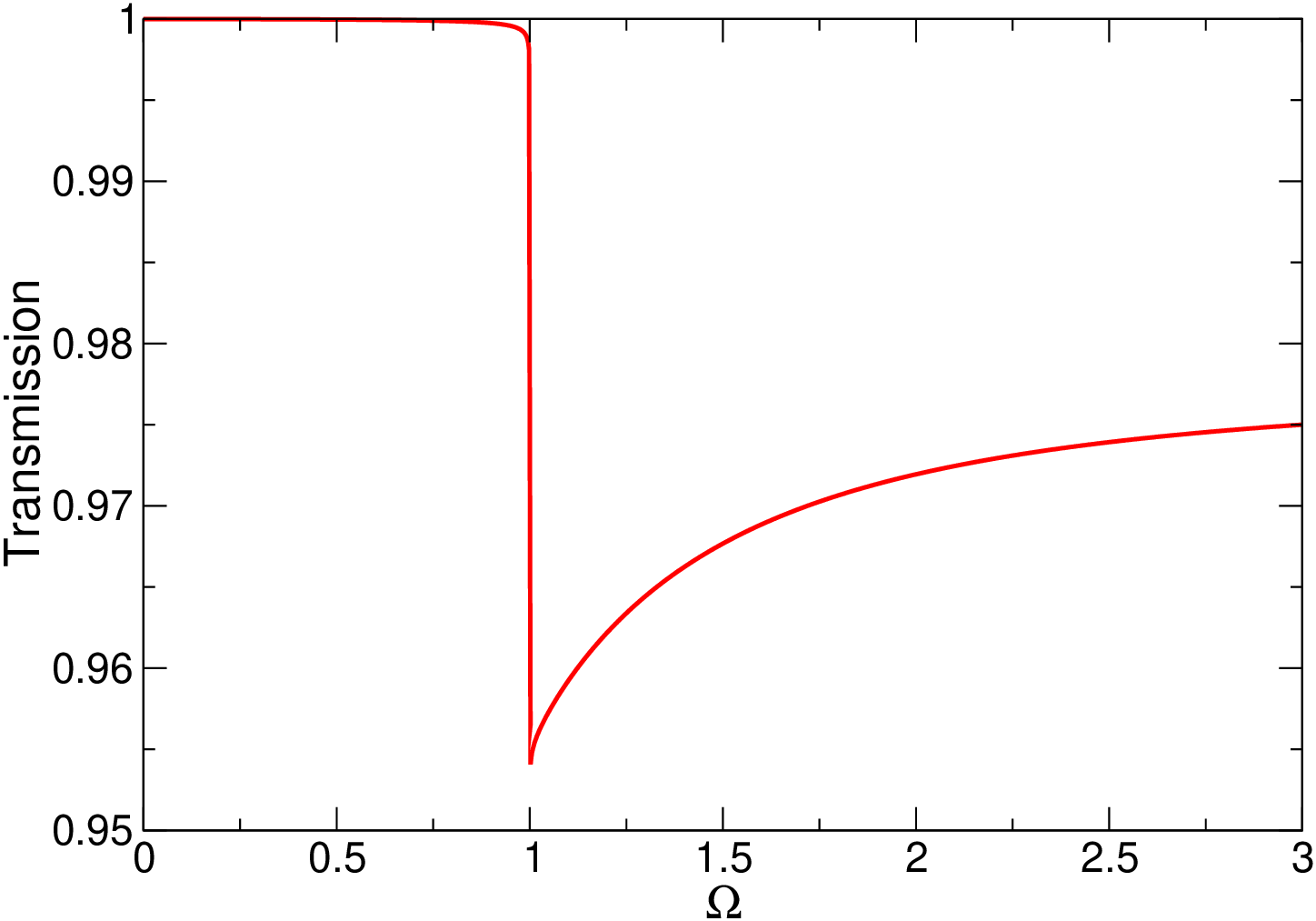}
\caption{\label{fig:T}The transmission coefficient of a wave passing through a 2D dielectric layer with the susceptibility $\chi^{(1),\rm{2D}}$, Eq. (\ref{chi1}), as a function of frequency $\Omega=\hbar\omega/2\Delta$.}
\end{figure}

If an electromagnetic wave is normally incident on a 2D dielectric layer characterized by the susceptibility (\ref{chi1}) the transmission coefficient is determined by the formula $T(\omega)=|1-2\pi i\omega\chi^{(1),\rm{2D}}(\omega)/c|^{-2}$. Its frequency dependence is shown in Figure \ref{fig:T}. At small frequencies, $\omega\ll c\Delta/e^2$, the function $T(\omega)$ decreases quadratically with the growing frequency,
\be 
T(\omega)=\left[1+\left(\frac{2e^2\omega}{3c\Delta}\right)^2\right]^{-2}\simeq 1-D\omega^2,
\ee
with the coefficient $D\propto 1/\Delta^2$ determined by the band gap. At frequencies higher that $2\Delta$ the inter-band absorption is switched on and the transmission coefficient falls down. Notice that at the absorption edge $\hbar\omega\simeq 2\Delta$ a single dielectric 2D layer (e.g., a monolayer of BN) absorbs more than 4\% of the incident radiation energy which is about twice as large as in graphene ($\simeq 2.3$\%).  

\begin{figure}%[t]
\includegraphics[width=0.98\columnwidth]{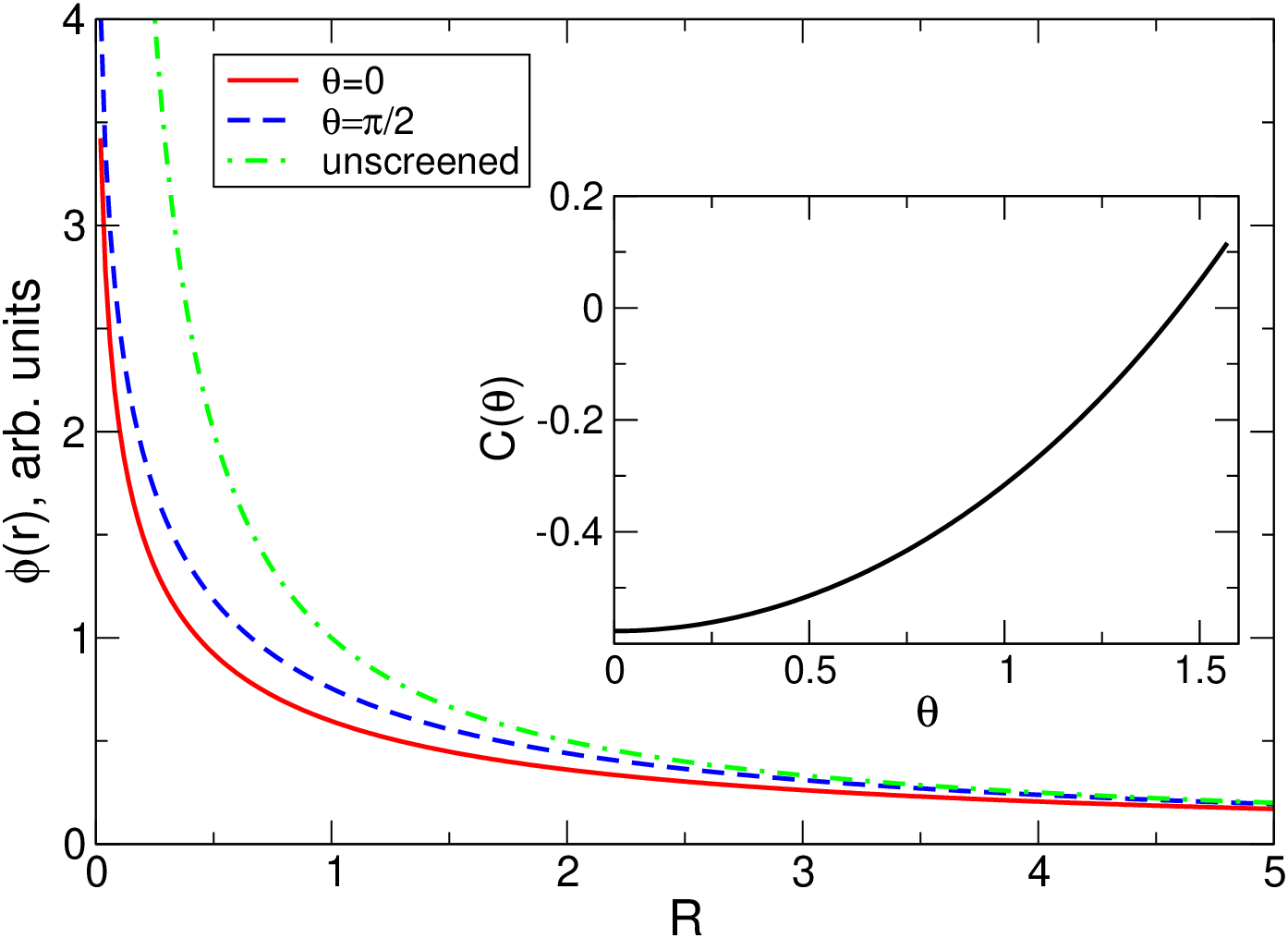}\\
\caption{\label{fig:scrpot}The potential (\ref{CoulSol}) as a function of $R=r/2\pi \chi^{(1),\rm{2D}}$ for $\theta=0$ (the direction normal to the 2D plane) and $\theta=\pi/2$ (the direction parallel to the 2D plane). The green dash-dotted curve shows the unscreened Coulomb potential $1/r$. Inset shows the function $C(\theta)$, Eq. (\ref{const}), in the interval from $\theta=0$ to $\theta=\pi/2$.}
\end{figure}

Now consider the static screening of a point charge $Q\delta(\textbf{r})=Q\delta(\textbf{r}_\parallel)\delta(z)$ placed in the plane of a 2D dielectric. The Poisson equation for the electric potential has the form 
\be 
\Delta \phi+4\pi \chi^{(1),\rm{2D}} \Delta_2\phi\delta(z)
=-4\pi Q\delta(\textbf{r}_\parallel)\delta(z),
\ee
Its solution,
\be 
\phi(\textbf{r})=
\frac Q{2\pi \chi^{(1),\rm{2D}}}\int_0^\infty \frac{f(\xi R,\theta)}{1+\xi}d\xi  ,
\label{CoulSol}
\ee
is shown in Figure \ref{fig:scrpot}; here $f(t,\theta)=e^{-t\cos\theta}J_0(t\sin\theta)$, $J_0$ is the Bessel function, $R=r/2\pi \chi^{(1),\rm{2D}}$ and $\theta$ is the angle between the vector $\textbf{r}$ and the $z$-axis. At a large distance from the charge, $R\gg 1$, Eq. (\ref{CoulSol}) gives the unscreened Coulomb potential $\phi(\textbf{r})\approx Q/r$. At smaller distances $R\lesssim 1$ the $r$-dependence of the screened Coulomb potential is logarithmic, 
\be 
\phi(r) \approx \frac Q{2\pi\chi^{(1),\rm{2D}}}\left(\ln\frac{2\pi\chi^{(1),\rm{2D}}}{r}+C(\theta)\right), 
\ee
and depends on the angle $\theta$,  
\be 
C(\theta)=\int_1^\infty \frac{f(t,\theta)}tdt
-\int_0^1 \frac{1-f(t,\theta)}tdt\label{const}.
\ee
In the directions perpendicular ($\theta=0$) and parallel ($\theta=\pi/2$) to the 2D plane the integrals in (\ref{const}) are calculated analytically, $C(0)=-\gamma$ and $C(\pi/2)=\ln 2-\gamma$ ($\gamma=0.577\dots$ is the Euler constant). At arbitrary angles $\theta$ the function $C(\theta)$ is shown in the Inset to Figure \ref{fig:scrpot}. The 2D dielectric substantially screens the field of the point charge at the distance smaller than or of order of the susceptibility length $\chi^{(1),\rm{2D}}$ which lies between $\sim 0.1$ $\mu$m and $\sim 0.1$ nm if the band gap varies from several meV to several eV. 

To summarize, we have shown that the standard procedure of averaging microscopic electromagnetic fields which is admitted in the macroscopic electrodynamics of 3D materials is inapplicable to 2D crystals like graphene and graphene related materials. We have performed a proper averaging of microscopic Maxwell equations and derived the corresponding macroscopic electrodynamic equations suitable for the description of 2D crystals. We have shown that electrodynamic properties of such crystals are adequately described by \textit{two-dimensional} quantities $\sigma^{{\rm 2D}}$ or $\chi^{{\rm 2D}}$. The three-dimensional quantities such as $\chi^{(3);{\rm 3D}}$, the dielectric function $\epsilon$, the refractive index $n=\sqrt{\epsilon}$ (linear and nonlinear) cannot be properly defined and should not be used in the electrodynamics of 2D crystals. 

The work has received funding from the European Union's Horizon 2020 research and innovation programme Graphene Core 2 under Grant Agreement No. 785219.

%\bibliographystyle{apsrev}  
%\bibliography{../../../../BIB-FILES/lowD,../../../../BIB-FILES/mikhailov,../../../../BIB-FILES/graphene}

%merlin.mbs apsrev4-1.bst 2010-07-25 4.21a (PWD, AO, DPC) hacked
%Control: key (0)
%Control: author (8) initials jnrlst
%Control: editor formatted (1) identically to author
%Control: production of article title (-1) disabled
%Control: page (0) single
%Control: year (1) truncated
%Control: production of eprint (0) enabled
%

\end{document}